\documentclass[page-classic]{epl2} 

\title{Impalement transitions in droplets impacting microstructured superhydrophobic surfaces}
\shorttitle{Impalement transitions in droplets impacting superhydrophobic surfaces} %Insert here a short version of the title if it exceeds 70 characters

\author{J. Hyv\"aluoma \and J. Timonen}
\shortauthor{J. Hyv\"aluoma \etal}

\institute{                    
  Department of Physics, University of Jyv\"askyl\"a, FI-40014 Jyv\"askyl\"a, Finland
}
\pacs{47.55.dr}{Interactions with surfaces}
\pacs{47.55.D-}{Drops and bubbles}
\pacs{68.08.Bc}{Wetting}

\abstract{
Liquid droplets impacting a superhydrophobic surface decorated with micro-scale
posts often bounce off the surface. However, by decreasing the impact velocity
droplets may land on the surface in a fakir state, and by increasing it posts may
impale droplets that are then stuck on the surface. We use a two-phase
lattice-Boltzmann model to simulate droplet impact on superhydrophobic surfaces,
and show that it may result in a fakir state also for reasonable
high impact velocities. This happens more easily if the surface is made more
hydrophobic or the post height is increased, thereby making the impaled
state energetically less favourable. 
}

\begin{document}

\maketitle

\section{Introduction} \label{sec:intro}

Recently, there has been rapidly growing interest in designing artificial 
surfaces with extreme hydrophobic properties. These efforts have been inspired 
in particular by biological superhydrophobic surfaces such as plant 
leaves or insect wings \cite{Neinhuis97,Lee04}. These surfaces have a hydrophobic 
coating and they typically have micron-scale roughness which has been found to
further enhance the hydrophobicity of the surface. Similar designs have been 
utilized in order to manufacture artificial, strongly water-repellent surfaces 
(see, e.g.,  refs.~\cite{Oner00,Lau03}).

The fact that roughness has a major effect on the hydrophobicity of a surface
has long been recognized. As a droplet is deposited on a rough hydrophobic surface, 
it can be found in two different states. In the first one, known as the impaled
or Wenzel state \cite{Wenzel36}, the droplet follows the surface topography and
no air is trapped beneath it. The second state is known as the fakir  
or Cassie-Baxter state \cite{Cassie44}. In this state the droplet sits on top 
of the roughnesses, and some air remains trapped in the hollows and grooves under 
the droplet. For the fakir droplets, a high contact angle is found. Also the 
contact angle of a droplet in the Wenzel state is higher than that observed for 
one on a smooth surface, although not as high as in the fakir state
\cite{Callies05}. Usually, a high contact angle is related to a low contact angle 
hysteresis, which is the most important property of a superhydrophobic surface, 
even though this relation is not generally true \cite{Oner00,Kiuru04,Hyvaluoma07}.
Therefore, in order to obtain strongly water-repellent behaviour,
surface designs leading to fakir droplets are preferred.

The fakir state is often metastable and therefore an impalement transition
to the Wenzel state may occur as a result of some external disturbance 
\cite{Journet05,Moulinet07,Reyssat08}. As the fakir state is typically 
the preferred one, it would be important to understand the robustness of this 
state, and the mechanisms leading to collapse to the Wenzel state, which may dramatically
change the hydrophobicity. This issue has recently attracted a lot of interest, 
and both experimental \cite{Moulinet07,He03,Pirat08} and numerical 
\cite{Pirat08,Dupuis05,Kusumaatmaja08} techniques have been utilized to elucidate 
the transition between different superhydrophobic states. These studies have 
been conducted on microstructured surfaces decorated with regular patterns 
of posts. The impalement transition can be triggered by several mechanisms, including,
among others, external pressure \cite{Journet05}, defects on the surface \cite{Moulinet07}, 
or hydrophilic contaminants deposited on the surface from the air or from the droplet
itself \cite{Reyssat08}.

One possibility to study impalement transitions is to consider droplets 
impacting microstructured hydrophobic surfaces \cite{Bartolo06,Reyssat06}.
In this case the dynamic `pressure' due to the impact velocity of the droplet
is the mechanism triggering the impalement transition. Recently, Bartolo
and co-workers \cite{Bartolo06} reported of 
three distinct regimes, in which the qualitative behaviour of an impacting 
droplet is different depending on the impact velocity. First, if the 
impact velocity is small enough, droplet lands on the posts and after 
some oscillations stays on the surface as a fakir droplet. Second, for 
intermediate impact velocities, droplet bounces off the surface.
Third, for high impact velocities, sticky droplets are observed. In
this regime, the posts on the surface impale the droplet, and liquid
penetrates the volume between the posts. Thus, in the last case,
the droplet is found in the Wenzel state.

In this work, we study droplets impacting on a hydrophobic surface patterned 
by posts with square cross section. We are especially interested in the
circumstances under which a droplet is impaled by posts, and when a
non-bouncing droplet is observed. Our approach is numerical and utilizes the 
lattice-Boltzmann method to simulate the behaviour of impacting 
droplets. In particular, we find the same three regimes as reported by 
Bartolo et al. in ref.~\cite{Bartolo06}. In addition, under certain circumstances, 
we find a fourth possible regime for impact velocities larger than those for 
bouncing droplets but smaller than the ones for sticky droplets. In this
state, a non-bouncing droplet that ends up in the fakir state is observed.
 
\section{Lattice-Boltzmann method and the Shan-Chen model} \label{sec:lb}

The present simulations were done using the lattice-Boltzmann (LB)
method. As this method is well established and throughly covered in
a number of review articles and books (see, e.g., 
refs.~\cite{Benzi92,Chen98,Wolf00,Succi01,Sukop06}), we only briefly discuss 
here, for the sake of completeness, the basics of the method.

In the LB method, the fluid is described by an ensemble
of particles moving along links between lattice nodes of a regular lattice.
Time and velocities are discretized such that during one time step particles
can move only to neighbouring lattice nodes. The LB fluid properties are 
determined by single-particle distribution functions $f_i({\bf r},t)$ which can be 
interpreted as the probabilities to find a particle at lattice node ${\bf r}$ 
at time $t$ moving with a discrete velocity ${\bf c}_i$. These distribution 
functions evolve according to the LB equation,
\begin{equation}
f_i({\bf r}+{\bf c}_i,t+1) - f_i({\bf r}_i,t) = 
-\frac{1}{\tau}\left( f_i({\bf r},t) - f_i^{eq}({\bf r},t) \right).
\end{equation}
Here, the right-hand side of the equation describes the collisions among the particles
as a relaxation process towards a local equilibrium which is a low-Mach-number
expansion of the Maxwell-Boltzmann distribution. We use the relaxation-time
approximation with a single characteristic time scale $\tau$.

The macroscopic quantities are obtained as velocity moments of the distribution
functions. In particular, the mass and momentum densities are given by
\begin{eqnarray}
\rho & = & \sum_i f_i \\
\rho {\bf u} & = & \sum_i {\bf c}_i f_i
\end{eqnarray}
respectively.

In order to model a two-phase fluid, we use the multiphase model developed
by Shan and Chen \cite{Shan93}. In this model a mean-field interparticle 
interaction is added to the standard LB equation. This force can be expressed
in the form
\begin{equation}
{\bf F} = G_b \psi({\bf r}) \sum_i t_i
          \psi({\bf r} + {\bf c}_i) {\bf c}_i,
\end{equation}
where $\psi = 1-\exp{(-\rho/\rho_0)}$ is an effective mass ($\rho_0$ is a 
reference density) and $G_b$ is a parameter that adjusts the strength of the 
interaction. The values of weight factors $t_i$ depend on the magnitude of 
the corresponding discrete velocity ${\bf c}_i$. Here we use the standard 
D3Q19 model (three-dimensional lattice with 19 discrete velocities), and the 
weights for this model can be found in ref.~\cite{Qian92}. This additional 
interaction enables simulation of liquid-vapour systems with surface tension.
The wetting behaviour of the liquid at solid walls is modelled with a similar 
type of force added between the solid and the fluid. Here we set a density value 
$\rho_w$ to the solid lattice nodes, and the contact angle can be modified 
through this density. 

\section{Results and discussion} \label{sec:results}

We use the following geometrical setup in the simulations. The size of the
system is $150 \times 150 \times 170$ lattice nodes, where the last dimension
is related to the vertical direction. To the bottom of the
system we add an array of posts with a square cross section of size 
$3 \times 3$. The distance between neighbouring posts is 10 lattice spacings. 
The length of the posts is between 10 and 25 lattice spacings.

In the middle of the simulation domain a droplet with a diameter
of 105 lattice spacings is initialized. In the beginning of a simulation,
we give the droplet an initial (impact) velocity $U_0$. We choose such values 
for the  simulation parameters that the density ratio of the liquid and 
vapour phases is approximately 30. Contact angle is slightly varied in 
the simulations, and the intrinsic contact angle (i.e., the contact angle
of a droplet on a perfectly smooth surface) is between 
$106^\circ$ and $120^\circ$. Notice that all lengths above were given in dimensionless
lattice units and the results below are given in terms of appropriate dimensionless
numbers.

\begin{figure}
\resizebox{0.95\columnwidth}{!}{
  \includegraphics{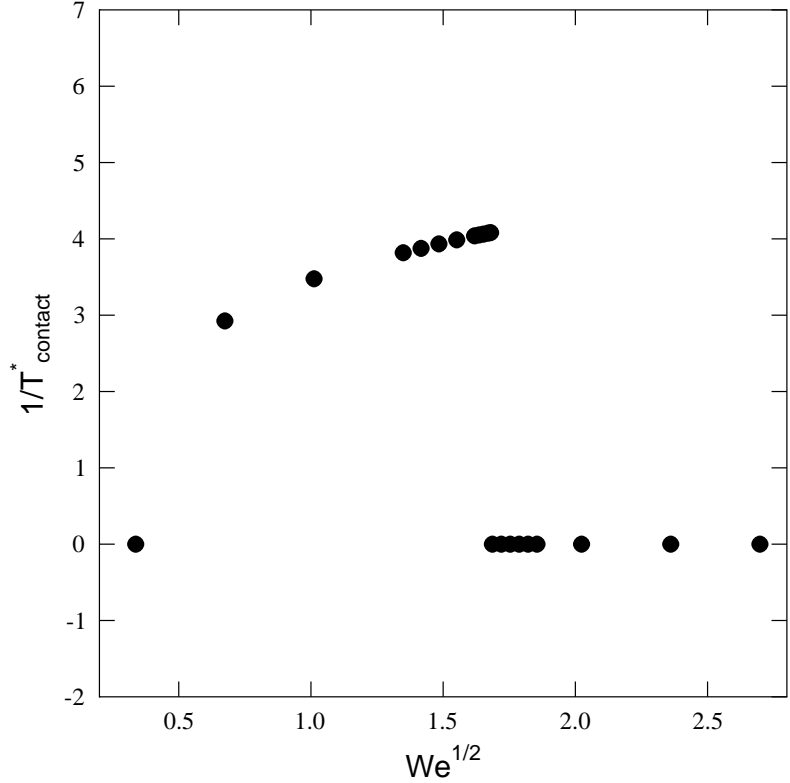}
}
\caption{The inverse of dimensionless contact time as a function $We^{1/2} = U_0 \sqrt{\rho R/\gamma}$
(impact velocity). Time is made dimensionless by scaling it by $t_0 = \rho R^2/\mu$ where
$\mu$ is the dynamic viscosity. The intrinsic contact angle is $\theta = 113^\circ$.
}
\label{fig:contact}
\end{figure}

First we simulated such a case in which the intrinsic contact angle 
was $113^\circ$, and the post height was 20 lattice spacings. We varied 
the initial velocity and measured the contact time,
i.e., the time that the droplet was in contact with the surface before bouncing
off. In the non-bouncing cases the contact time obviously approaches infinity.
Experimental studies \cite{Bartolo06,Richard02} and scaling arguments
\cite{Richard02} suggest that contact time does not depend on the impact
velocity if this velocity is high enough. However, in their experiments 
Richard et al. found that, when the Weber number (the ratio of fluid
inertia to surface tension) was $We = \rho U_0^2 R/\gamma \ll 1$, contact time 
increased with decreasing impact velocity \cite{Richard02}. With the parameters 
used in the present simulations, 
$We$ varied between $0.1$ and $7$. Therefore, the contact times shown in 
fig.~\ref{fig:contact} behave as expected. For high impact velocities 
the contact time stays essentially constant, but for somewhat lower  
velocities, when $We$ is decreased below unity, it increases for decreasing 
impact velocity.

\begin{figure}
\resizebox{0.8\columnwidth}{!}{
  \includegraphics{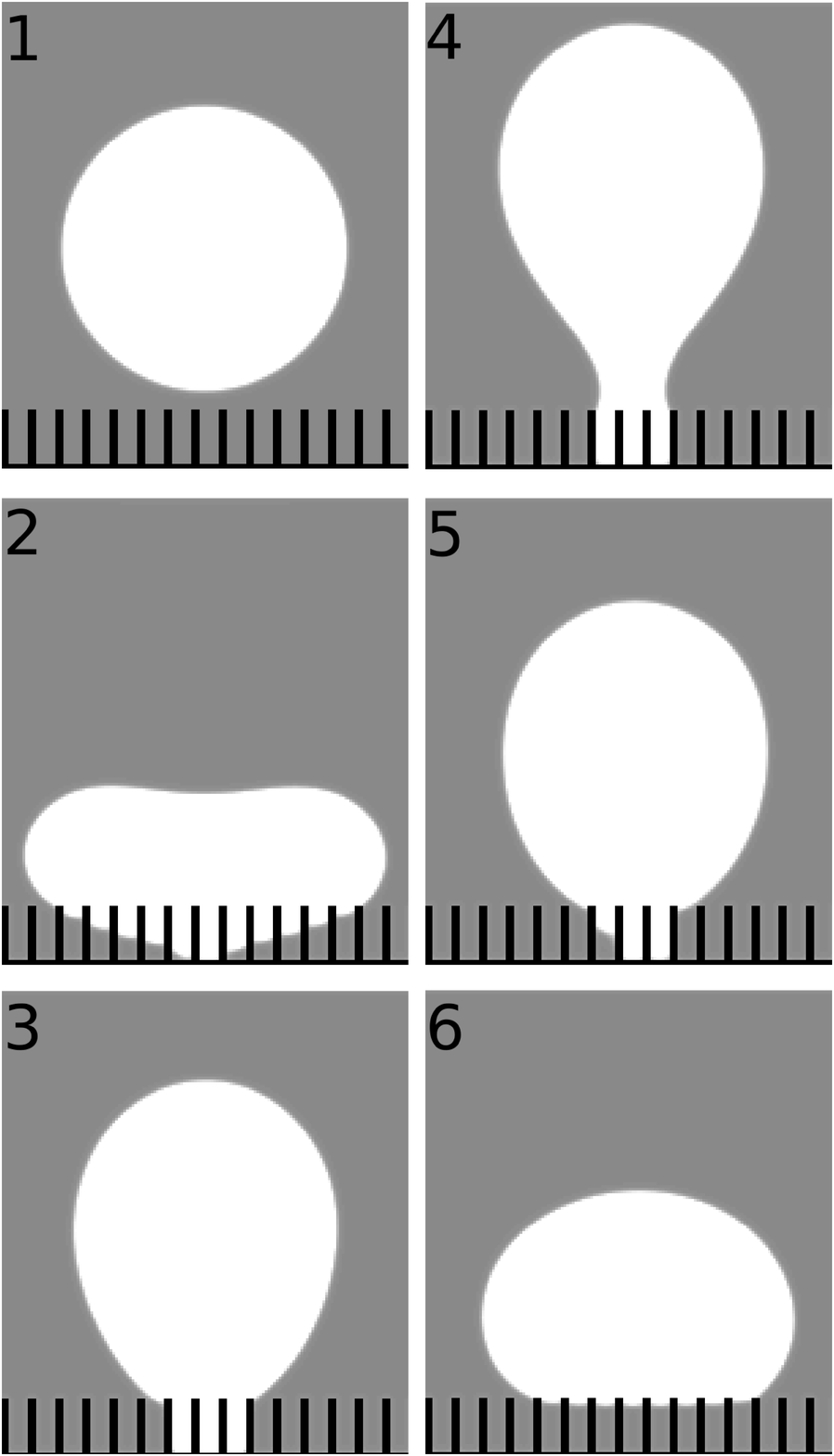}
}
\caption{Cross sections of a 3D droplet-impact simulation in the second
non-bouncing regime. 
(1) The initial droplet. (2) Droplet impacts the
surface and starts to fill the volume between the posts. (3) Droplet 
tries to bounce off but remains stuck at the centre of the contact area. (4)
Maximum height for the centre of mass during the attempted bouncing. (5) 
Centre-of-mass motion is again downwards but, at the same time, the volume
between posts begins to be drained. (6) Droplet in its
final state.
}
\label{fig:snapshots}
\end{figure}

Superficially the contact times shown in fig.~\ref{fig:contact} 
lead us to conclude that there are three different regimes in the qualitative
behaviour of droplet impact. First, when the impact velocity is low,
the contact time is infinite. Second, there is a regime of an almost
constant contact time. Third, another regime with infinite contact
time is observed at high impact velocities. These regimes would
correspond to the non-bouncing, bouncing, and sticky
droplets, respectively, as in ref.~\cite{Bartolo06}. However, a 
closer inspection of the
last regime reveals two different types of qualitative behaviour.
For very high impact velocities we find the sticky behaviour
where the posts impale the droplet. But for lower impact
velocities when the contact time still diverges, the droplet 
is eventually found in the fakir state.
In fig.~\ref{fig:snapshots} we show a series of snapshots
from the impact process in this case. It is evident that posts
impale the droplet, but as the droplet bounces, the interpost
volume is drained during the process. However, the droplet does
not bounce off but remains on the surface and finally 
ends up in the fakir state. We thus find it in a similar
state as in the non-bouncing case although the two processes 
are different. We distinguish between these two states
by calling them the first and second non-bouncing state corresponding
to an impact velocity that is lower and respectively higher than that 
of the bouncing droplet.

\begin{figure}
\resizebox{0.95\columnwidth}{!}{
  \includegraphics{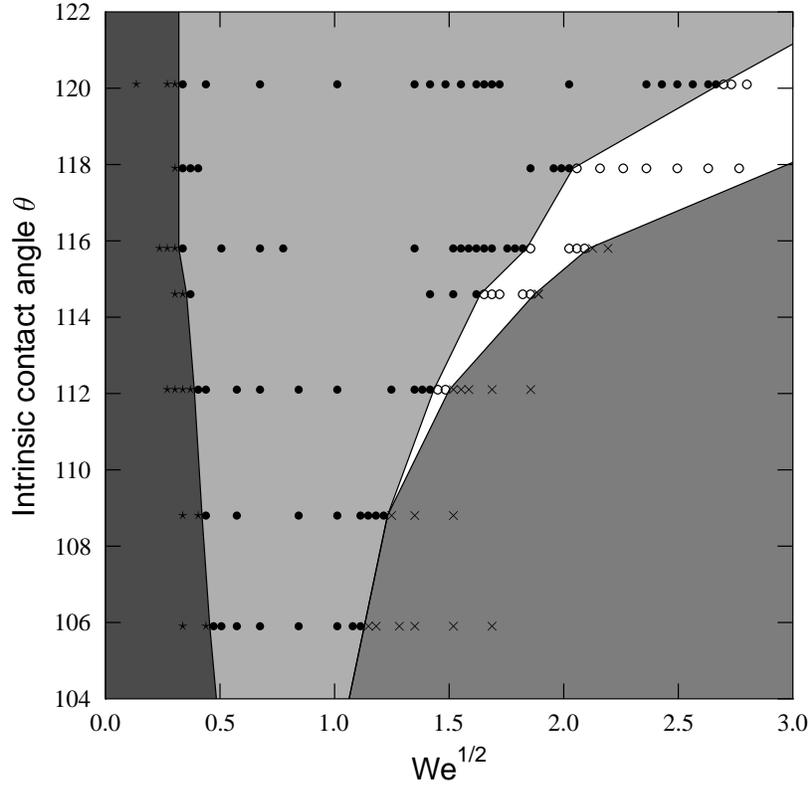}
}
\caption{
Impact regimes for varying hydrophobicity (intrinsic contact angle)
and impact velocity ($\sqrt{We} = U_0 \sqrt{\rho R/\gamma}$). The regimes are first non-bouncing,
bouncing, second non-bouncing (white), and sticky regime, from left to right.
The simulation results are depicted with stars, filled circles, open
circles, and crosses, respectively. The post height in all simulations
was 15 lattice spacings.
}
\label{fig:phase1}
\end{figure}

\begin{figure}
\resizebox{0.95\columnwidth}{!}{
  \includegraphics{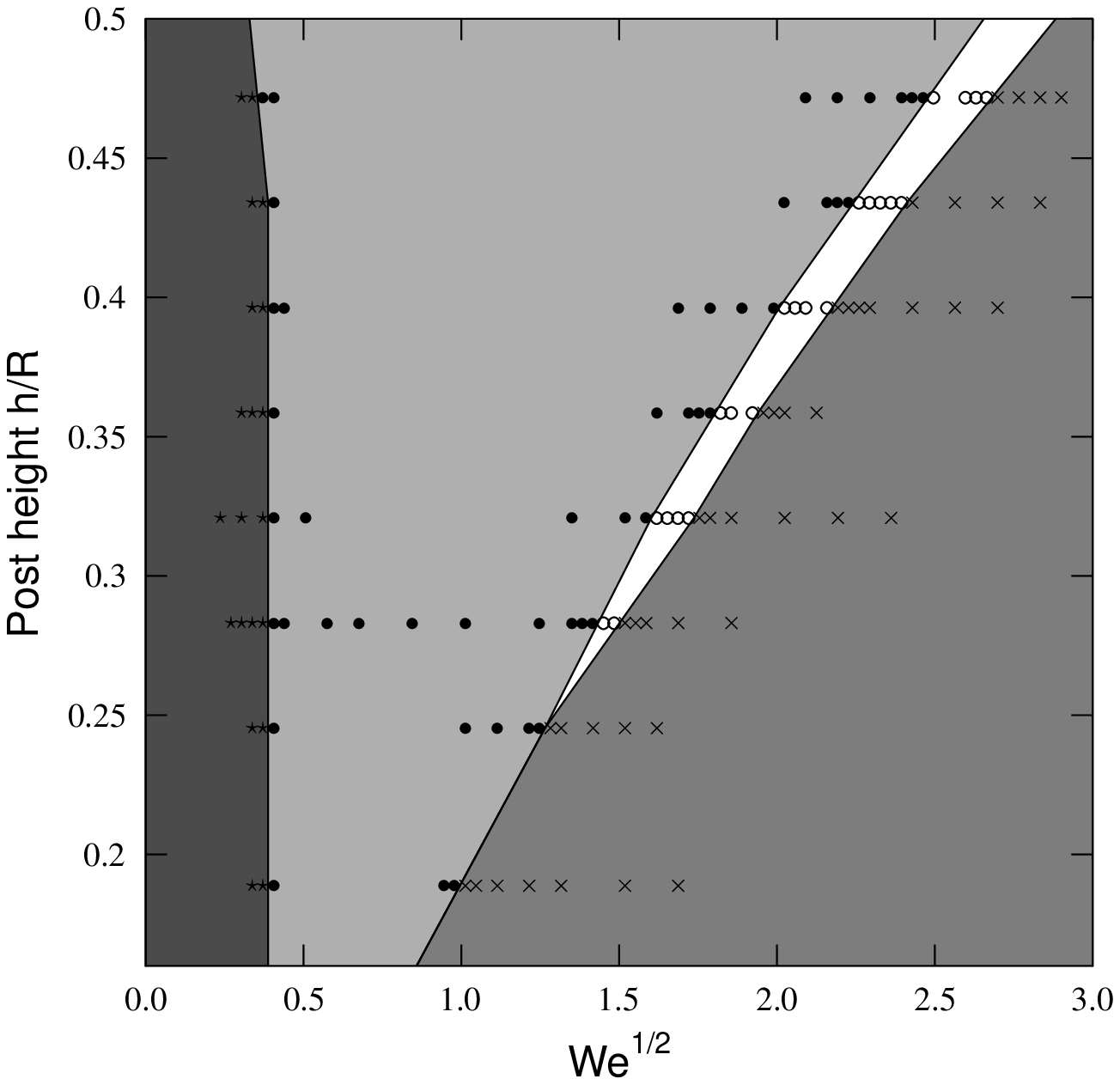}
}
\caption{
Impact regimes for varying post height and impact velocity ($\sqrt{We} = U_0 \sqrt{\rho R/\gamma}$).
The post height is made dimensionless by scaling it by droplet radius $R$. 
The intrinsic contact angle in all simulations was $112^\circ$.
The colours and symbols are as in fig. \ref{fig:phase1}. 
}
\label{fig:phase2}
\end{figure}

Next we considered how the hydrophobicity of the surface affects the
occurrence of the second non-bouncing state. To this end, we simulated
droplet impact on surfaces with varying hydrophobicity such that the intrinsic
contact angle varied between $106^\circ$ and
$120^\circ$. The post heigth was fixed to 15 lattice spacings and 
also the impact velocity was varied so as to
find the velocity intervals where different impact regimes, and especially
the second non-bouncing state, are found. The results are shown in fig. 
\ref{fig:phase1}. We observe that, as the hydrophobicity grows (i.e., the 
contact angle increases), the second non-bouncing state is found
at higher impact velocities, but also the velocity interval
of this state becomes wider. 
Also notice that, for the lowest values of contact angle 
used in the simulations, we do not observe
the second non-bouncing state at all regardless of the impact velocity.
It is thus evident that 
the second non-bouncing state is found easier if the surface
is made more hydrophobic.

In a similar fashion, we studied the effect of post height.
The contact angle was kept contant, $\theta = 112^\circ$,
and the simulations were performed on surfaces with post heights 
between 10 and 25 lattice spacings. The results of these
simulations are shown 
in fig. \ref{fig:phase2}. The qualitative observations from these 
simulations are similar to those reported above. As the height of the 
posts is increased, a wider interval of impact velocities leads to
the second non-bouncing state. Also, a higher impact velocity is
necessary for this state as the post height increases,
and when the posts are short enough, the second non-bouncing state
is absent.

The results described above are related to the metastability
of fakir droplets. As already discussed in the Introduction,
the fakir state is typically metastable, but transition
to the energetically more favourable impaled state requires
external triggering in order to overcome the energy barrier
separating these two states. Our investigation shows what happens
to an impacting droplet when the impaled state is made less and 
less favourable. This is
exactly what happens when the hydrophobicity of the surface or the
length of the posts are increased. 

The intrinsic contact angles leading to the second non-bouncing state 
appears to be quite large. However, such values
are still experimentally achievable as the largest contact angle 
observed on a smooth surface is presumably about $120^\circ$ as reported by Nishino 
and co-workers \cite{Nishino99}. These authors used trifluoromethyl
carbon self-assembled on a surface. Another possibility to achieve
high contact angles could be to utilize roughness at different length
scales such that the patterned surface itself would be made rough with
some smaller-scale roughness. This type of surface design has been proposed
by Patankar \cite{Patankar04}.

\section{Summary} \label{sec:summary}

In conclusion, we simulated droplets impacting hydrophobic surfaces patterned
with regular arrays of posts. Our results were in agreement with those of Bartolo 
et al. \cite{Bartolo06}, i.e., the behaviour of an impacting droplet
depends on the impact velocity such that a non-bouncing, bouncing or sticky
droplet results from the impact as the impact velocity increases.
We found in addition that, when the impaled state is made energetically
less favourable, another non-bouncing state is observed between
the bouncing and sticky state.
This state might be useful for gaining better
understanding of the energy-barrier related mechanisms in
droplet impalement and thus for the development of more
robust superhydrophobic surfaces.

\end{document}